\documentclass[12pt]{iopart}

\usepackage{iopams}  

\usepackage{bm}
\usepackage{xcolor}
\usepackage{mathrsfs}
\usepackage[utf8]{inputenc}
\usepackage{enumerate}
\usepackage{appendix}
\usepackage{dcolumn}
\usepackage{multirow}
\usepackage{float}

\begin{document}

\title[Generalized Rainich conditions]{Generalized Rainich conditions, generalized stress-energy conditions,
and the Hawking--Ellis classification}

\author{Prado Mart\'{\i}n--Moruno$^1$ and Matt Visser$^2$}

\address{$^1$Departamento de F\'isica Te\'orica I, Universidad Complutense de Madrid, \\E-28040 Madrid, Spain}
\address{$^2$School of Mathematics and Statistics, Victoria University of Wellington, \\ PO Box 600, Wellington 6140, New Zealand}
\ead{pradomm@ucm.es, matt.visser@sms.vuw.ac.nz}
\vspace{10pt}
\begin{indented}
\item[]\today
\end{indented}

\begin{abstract}
The (generalized) Rainich conditions are \emph{algebraic} conditions which are polynomial in the (mixed-component) stress-energy tensor. As such they are logically distinct from the usual classical energy conditions (NEC, WEC, SEC, DEC), and logically distinct from the usual Hawking--Ellis (Segr\'e--Pleba\'nski) classification of stress-energy tensors (type I, type II, type III, type IV). 
There will of course be significant inter-connections between these classification schemes, which we explore in the current article. 
Overall, we shall argue that it is best to view the  (generalized) Rainich conditions as a refinement of the classical energy conditions and the usual Hawking--Ellis classification. 
\end{abstract}


\vspace{2pc}
characteristic polynomial, minimal polynomial.

\maketitle

\def\tr{{\mathrm{tr}}}
\def\cof{{\mathrm{cof}}}
\def\pdet{{\mathrm{pdet}}}

%

\section{Introduction}

The usual classical energy conditions, (NEC, WEC, SEC, DEC, and their variants), are most typically used within the context of various singularity theorems in general relativity, where they are used to enforce focussing (or defocussing) of null or timelike geodesics~\cite{Hell,Martin-Moruno:2013a,Martin-Moruno:2013b,Martin-Moruno:2015,LNP,Visser:1994,Visser:1997, Balakrishnan:2017,Fu:2017,Fu:2016, Koeller:2015, Bousso:2015}. Similarly the  usual Hawking--Ellis (Segr\'e--Pleba\'nski) classification of stress-energy tensors, (type I, type II, type III, type IV), is also most typically used in special and general relativity, wherein this Hawking--Ellis classification effectively controls the extent to which the stress-energy tensor can be diagonalized by local Lorentz transformations~\cite{Hell,Plebanski:1964,LNP}. (It is the Lorentzian signature of spacetime that makes this non-trivial.) 

Complementing and refining these two classification schemes we shall develop a version of the (generalized) Rainich conditions. The usual Rainich condition amounts to the observation that the (mixed-component) stress-energy tensor of the classical electro-magnetic field $T^a{}_a$ satisfies the purely \emph{algebraic} constraints~\cite{1925,Misner:1957,Witten:1959}
\begin{equation}
(T^2)^a{}_b =  {\tr(T^2)\over4} \; \delta^a{}_b;  \qquad   \tr(T)=0.
\end{equation}
(See appendix \ref{A:B} for a sketch of a proof.)
When inserted into the Einstein equations this implies that any \emph{electro-vac} spacetime can be (partially) characterized by simple \emph{purely geometric} statements regarding the (mixed-component) Ricci tensor $R^a{}_b$:
\begin{equation}
(R^2)^a{}_b =  {\tr(R^2)\over4} \; \delta^a{}_b;  \qquad   \tr(R)=0.
\end{equation}
This is the mathematical basis of the so-called ``already unified'' approach to the long sought for unification of classical gravity and classical electromagnetism. We shall seek to generalize this observation as much as possible, somewhat along the lines of references~\cite{Senovilla:2000, Bergqvist:2001,Senovilla:2002,Bergqvist:2004,Plebanski:1994,Torre:2013,Krongos:2015,Santos:2016,Balfagon:2007}. We will be working within classical general relativity, aiming for algebraic constraints on the stress-energy tensor and Ricci tensor that can be related to simple physical statements regarding the material sources.

The main technical tools we will use are based on considerations of the generalized eigenvalue problem
\begin{equation}
(T_{ab}-\lambda \,g_{ab})  V^b =0,
\end{equation}
which we recast as
\begin{equation}
(T^a{}_{b}-\lambda\, \delta^a{}_{b})  V^b =0.
\end{equation}
It is the observation that the mixed-component $T^a{}_{b}$ is \emph{not} symmetric that is the source of all the technical difficulties.\footnote{In Euclidean (4+0) signature everything trivializes and all stress energy tensors are type I~\cite{LNP}. 
Physically, we are interested in Lorentzian (3+1) $\equiv$ (1+3) signature. We shall deem (2+2) signature physically inappropriate, though we shall sometimes encounter it in the mathematical analysis below.} The main mathematical tools we will use are general properties of matrix analysis, in particular the characteristic polynomial, the minimal polynomial, and the Jordan normal form~\cite{HJ1,HJ2}.

While the mixed-component matrix $T^a{}_{b}$ is certainly \emph{not} symmetric, it is also not the most general asymmetric matrix possible. Indeed,  in an orthonormal basis, $T^a{}_{b}$ is of the form
\begin{equation}
T^a{}_b = \left[ \begin{array}{c|c} \rho& f^j \\ \hline - f_i & \pi^{ij} \end{array} \right]
\end{equation}
with $\pi^{ij}$ being symmetric. (Algebraically this corresponds, in an orthonormal basis, to the mixed tensor $T^a{}_{b}$ satisfying $T^\mathrm{transpose} = \eta\,T\,\eta$.)
 Because of this algebraic structure (and avoiding interchange of columns and rows) not all Jordan normal forms need necessarily arise, (and the interplay between Jordan normal forms and the timelike/null/spacelike nature of the eigenvectors is nontrivial).

This paper can be outlined as follows: Section \ref{sec:main} contains the main body of the paper. In section \ref{sec:f} we present the general mathematical framework on which we base the new stress-energy tensor classification. In section \ref{sec:ps} we discuss this classification in detail, emphasizing the physically interesting cases. We then present some applications of this new classification; these are the formulation of some generalized Rainich conditions, presented in section \ref{sec:Rc}, and the relation between various energy conditions, explicated in section \ref{sec:EC}. In section \ref{sec:d} we discuss our results. Finally, we include some comments about the Hawking--Ellis classification in appendix \ref{A:A}, summarize the classic Rainich algebraic conditions in appendix \ref{A:B}, and consider the classification for $(1+1)$-dimensional scenarios in appendix \ref{A:C}.

\section[Stress-energy  tensor classification]{Stress-energy tensor classification: \\Characteristic and minimal polynomials}\label{sec:main}
We shall classify stress-energy tensors using their \emph{algebraic} properties.
\subsection{Framework}\label{sec:f}
Let us consider the stress-energy tensor $T^{ab}$ and lower one index: $T^a{}_b$.  More formally, if the specific indices are not important, we write $T^\bullet{}_\bullet$.
One can now construct the Lorentz-invariant \emph{characteristic polynomial} 
\begin{equation}
c(\lambda) = \det\left( T^\bullet{}_\bullet - \lambda\, \delta^\bullet{}_\bullet \right).
\end{equation}
This can be written in terms of the (distinct) eigenvalues of  $T^a{}_b$ as
\begin{equation}
c(\lambda) = \prod_i (\lambda-\lambda_i)^{n_i}=\lambda^4+a_{3}\lambda^{3}+...+a_1\lambda+a_0,
\end{equation}
where $n_i$ is the (algebraic) multiplicity of the eigenvalue $\lambda_i$, with $\sum_i n_i=4$ in any 4-dimensional spacetime.
Furthermore, from the Cayley--Hamilton theorem we know
\begin{equation}
 c(T^\bullet{}_\bullet)=0.
\end{equation}
This implies in particular that (in 4 dimensions) the 4th power of the stress-energy tensor is a always cubic polynomial of lower powers
\begin{equation}
(T^4)^a{}_b =  p_3 (T^a{}_b) = \sum_{j=0}^3 k_i \, (T^i)^a{}_b,
\end{equation}
 where $(T^2)^a{}_b = T^a{}_c\, T^c{}_b$, $(T^3)^a{}_b = T^a{}_c\,T^c{}_d\, T^d{}_b$, and $(T^4)^a{}_b = T^a{}_c\,T^c{}_d\,T^d{}_f\, T^f{}_b$. 
This is the most general (and weakest) Rainich-like condition one might encounter, but it is more useful if one refines this condition with extra physical information.
For instance, the  \emph{minimal polynomial} for $ T^a{}_b$ is the lowest-degree polynomial $m(\lambda)$ such that $m( T^\bullet{}_\bullet) = 0$.
The degree of the minimal polynomial can, therefore, run from 1 to 4 in (3+1) dimensions.
We have
\begin{equation}
 m(\lambda) = \prod_i (\lambda-\lambda_i)^{m_i},
\end{equation}
where $m_i$ is the dimension of the largest Jordan block corresponding to eigenvalue $\lambda_i$. Hence we have $1\leq m_i\leq n_i$ and so $ 1\leq \sum_i 1 \leq \sum_i m_i \leq 4$. 

In view of this, we can consider a classification of stress-energy tensors according to the degree of their minimal polynomial $m(\lambda)$. This, in a (3+1)-dimensional spacetime we can have four different classes of stress-energy tensors, defined by having a minimal polynomial of degree 1, 2, 3 or 4, respectively. Each class will in turn be composed of different sub-classes of stress-energy tensors depending of the spectral decomposition of the matrix $T^\bullet{}_\bullet$. We shall discuss these cases in detail throughout the next sections, whereas we consider the $1+1$-dimensional case in appendix \ref{A:C}.

\subsection{Physical scenarios}\label{sec:ps}

Let us now explicitly write down the possible classes and sub-classes of stress-energy tensors according to this classification, giving some examples of relevant situations of physical interest that can be described by these stress-energy tensors.

\paragraph{Degree 1:} The only possibility is $m(\lambda) = (\lambda-\lambda_*)$ and $c(\lambda)=(\lambda-\lambda_*)^4$. So we have only one eigenvalue, which has to be real. That is:
\begin{equation}
T^a{}_b \sim \left[\begin{array}{cccc} 
\lambda_* &0 &0 &0\\ 0 &\lambda_*& 0 &0\\ 0&0& \lambda_* & 0\\0&0&0&\lambda_*\\\end{array}\right].
\end{equation}
This is a special case of type I according to the Hawking--Ellis classification, where $\lambda_* = -\rho = p_1 = p_2 = p_3$. 
Physically this describes \emph{vacuum energy}.


\paragraph{Degree 2:} There are two sub-cases:
\begin{description}
\item[I:] Only one distinct eigenvalue, which has to be real. So, $m(\lambda) = (\lambda-\lambda_*)^2$, and $c(\lambda)=(\lambda-\lambda_*)^4$. Then
\begin{equation}\label{12}
T^a{}_b \sim \left[\begin{array}{cc|cc} 
\lambda_* &1 &0 &0\\ 0 &\lambda_*& 0 &0\\ \hline 0&0& \lambda_* & 0\\0&0&0&\lambda_*\\\end{array}\right].
\end{equation}
This is type II in the special case that $\lambda_* = -\mu =p_1=p_2=p_3$. \\
(See appendix \ref{A:A} for conventions.)\\
Physically this corresponds to a \emph{null flux parallel to the x-axis superimposed on a EM field parallel to the x-axis}.

\item[II:] Two distinct eigenvalues, then $m(\lambda) = (\lambda-\lambda_1)(\lambda-\lambda_2)$. There are two sub-cases:
\begin{description}
\item[a:] 
$c(\lambda)=(\lambda-\lambda_1)^2(\lambda-\lambda_2)^2$. Then
\begin{equation}
T^a{}_b \sim \left[\begin{array}{cc|cc} 
\lambda_1 &0 &0 &0\\ 0 &\lambda_1& 0 &0\\ \hline 0&0& \lambda_2 & 0\\0&0&0&\lambda_2\\\end{array}\right],
\end{equation}
where both eigenvalues have to be real.\footnote{If the eigenvalues were to be complex they would have to be a repeated complex conjugate pair, but this is not compatible with (3+1) signature, it would imply (2+2) signature. To see this, rewrite $\lambda_2 = \lambda_1^*$, and rearrange $T^a{}_b$ to read
\begin{equation*}
T^a{}_b \sim \left[\begin{array}{cc|cc} 
\lambda_1 &0 &0 &0\\ 0 &\lambda_1^*& 0 &0\\ \hline 0&0& \lambda_1 & 0\\0&0&0&\lambda_1^*\\\end{array}\right].
\end{equation*}
Each of these two $2\times2$ blocks corresponds to (1+1) dimensional type IV, see \ref{A:A}, so the $4\times4$ matrix is only compatible with (2+2) signature.
} \\
This is special case of type I with $\lambda_1=-\rho=p_1$, and $\lambda_2 = p_2=p_3$. \\
Interesting physical examples are specific \emph{spherical symmetric} scenarios with $\rho=-p_r$~\cite{Jacobson:2007}, and  a \emph{non-null EM field}
when $\lambda_1=-\lambda_2$.

\item[b:]
$c(\lambda)=(\lambda-\lambda_1)(\lambda-\lambda_2)^3$. Then
\begin{equation}
\hspace{-1cm}
T^a{}_b \sim \left[\begin{array}{c|ccc} 
\lambda_1 &0 &0 &0\\ \hline 0 &\lambda_2& 0 &0\\ 0&0& \lambda_2 & 0\\0&0&0&\lambda_2\\\end{array}\right] \quad {\rm or}\quad 
T^a{}_b \sim \left[\begin{array}{ccc|c} 
\lambda_1 &0 &0 &0\\  0 &\lambda_1& 0 &0\\ 0&0& \lambda_1 & 0\\ \hline 0&0&0&\lambda_2\\\end{array}\right],
\end{equation}
with, of course, real eigenvalues. Note that, as in this case the Jordan form is diagonal, it is not important whether the triple eigenvalue is associated only to spacelike eigenvectors or to spacelike and a timelike eigenvector.

The stress-energy tensor on the left is a special case of type I with $\lambda_1=-\rho$, and $\lambda_2 = p_1=p_2=p_3$,
describing, for example, a \emph{perfect fluid} (if $\lambda_2=0$ this specializes to \emph{dust}).  This can also be used to describe a \emph{scalar field}.

The stress-energy tensor on the right is also a special case of type I, now  with $\lambda_1=-\rho=p_1=p_2$ and $\lambda_2 = p_3$.
When $p_3=3\rho$ this describes the  \emph{Casimir vacuum between parallel plates}.
\end{description}
\end{description}


\paragraph{Degree 3:} Here we have three possibilities:
\begin{description}
\item[I:] Only one distinct eigenvalue, which must be real. \\
We have $m(\lambda) = (\lambda-\lambda_*)^3$ and $c(\lambda)=(\lambda-\lambda_*)^4$.  Then
\begin{equation}\label{14}
T^a{}_b \sim \left[\begin{array}{ccc|c} 
\lambda_* &1 &0 &0\\ 0 &\lambda_*& 1 &0\\ 0&0& \lambda_* & 0\\ \hline0&0&0&\lambda_*\\\end{array}\right].
\end{equation}
This is a special case of type III with $\lambda_*=-\rho=p_3$. (See \ref{A:A}.)\\
This form of stress-energy tensor \emph{does not occur} classically in nature, and does not even seem to occur semi-classically.

\item[II:] Two distinct eigenvalues, then $m(\lambda) = (\lambda-\lambda_1)^2(\lambda-\lambda_2)$. There are two sub-cases.
\begin{description}
\item[a:] 
$c(\lambda)=(\lambda-\lambda_1)^2(\lambda-\lambda_2)^2$. Then\\
\begin{equation}\label{15}
T^a{}_b \sim \left[\begin{array}{cc|cc} 
\lambda_1 &1 &0 &0\\ 0 &\lambda_1& 0 &0\\ \hline 0&0& \lambda_2 & 0\\0&0&0&\lambda_2\\\end{array}\right].
\end{equation}
This is type II in the special case $\lambda_1 = -\mu$, and $\lambda_2 = p_1=p_2$.\\
Physically this corresponds, for example, to a  \emph{null flux superimposed on spherical or planar symmetry}.

\item[b:]
$c(\lambda)=(\lambda-\lambda_1)^3(\lambda-\lambda_2)$. Then\\
\begin{equation}\label{16}
T^a{}_b \sim \left[\begin{array}{cc|c|c} 
\lambda_1 &1 &0 &0\\ 0 &\lambda_1& 0 &0\\ \hline 0&0& \lambda_1 & 0\\ \hline 0&0&0&\lambda_2\\\end{array}\right].
\end{equation}
This is type II in the special case $\lambda_1 = -\mu=p_2$ and $\lambda_2 = p_3$.\\
Physically this corresponds, for example, to a \emph{null flux} superimposed on a somewhat specific background (with the quantity of the null flux degenerate with the amount of stress in one of the orthogonal spacelike directions).
\end{description}

\item[III:] Three distinct eigenvalues, then $m(\lambda) = (\lambda-\lambda_1)(\lambda-\lambda_2)(\lambda-\lambda_3)$. \\
Then $c(\lambda)=(\lambda-\lambda_1)^2(\lambda-\lambda_2)(\lambda-\lambda_3)$. So we have
\begin{equation}
\hspace{-2cm}
T^a{}_b \sim \left[\begin{array}{cc|cc} 
\lambda_1 &0 &0 &0\\ 0 &\lambda_1& 0 &0\\ \hline 0&0& \lambda_2 & 0\\0&0&0&\lambda_3\\ \end{array}\right]\, {\rm or}\,
\left[\begin{array}{c|cc|c} 
\lambda_2 &0 &0 &0\\ \hline 0 &\lambda_1& 0 &0\\  0&0& \lambda_1 & 0\\ \hline 0&0&0&\lambda_3\\ \end{array}\right]\,{\rm or}\,
\left[\begin{array}{cc|cc} 
\lambda_2 &0 &0 &0\\ 0 &\lambda_3& 0 &0\\ \hline 0&0& \lambda_1 & 0\\0&0&0&\lambda_1\\ \end{array}\right]
\end{equation}
Analogous to the situation in case degree 2IIb, it is not important whether the double eigenvalue is associated only to a pair of spacelike eigenvectors, or to a spacelike and a timelike eigenvector. 
From left to right, this is a specialization of type I, first with $\lambda_1=-\rho=p_1$, $\lambda_2=p_2$, and $\lambda_3=p_3$;
second with $\lambda_1=-\rho$, $\lambda_2=p_1=p_2$, and $\lambda_3=p_3$; and finally, with
\emph{spherical or planar symmetry}, $\lambda_2=-\rho$, $\lambda_3=p_1$, and $\lambda_1=p_2=p_3$. 

\end{description}

\paragraph{Degree 4:} There are now four possibilities:

\begin{description}
\item[I:] Only one distinct eigenvalue, then $m(\lambda) = (\lambda-\lambda_*)^4$ and $c(\lambda)=(\lambda-\lambda_*)^4$. We have
\begin{equation}\label{18}
T^a{}_b \sim \left[\begin{array}{cccc} 
\lambda_* &1 &0 &0\\ 0 &\lambda_*& 1 &0\\ 0&0& \lambda_* & 1\\ 0&0&0&\lambda_*\\\end{array}\right].
\end{equation}
\emph{It cannot exist}, since this (algebraic) case is not compatible with the Hawking--Ellis classification, and this incompatibility is ultimately due to the fact that this case is incompatible with (3+1) Lorentzian signature. Specifically, this particular case has no spacelike eigenvector, in contrast to all types in the Hawking--Ellis classification.

\item[II:] Two distinct eigenvalues. There are two sub-cases.
\begin{description}
\item[a:] 
$m(\lambda) = (\lambda-\lambda_1)^2(\lambda-\lambda_2)^2$ 
and $c(\lambda)=(\lambda-\lambda_1)^2(\lambda-\lambda_2)^2$. We have
\begin{equation}
T^a{}_b \sim \left[\begin{array}{cc|cc} 
\lambda_1 &1 &0 &0\\ 0 &\lambda_1& 0 &0\\ \hline 0&0& \lambda_2 & 1\\0&0&0&\lambda_2\\\end{array}\right].
\end{equation}
\emph{It cannot exist, at least not in (3+1) dimensions}.\footnote{The $4\times4$ matrix above block diagonalizes into two (1+1) dimensional type II stress-energy tensors,  so it corresponds to physically inappropriate (2+2) signature.}

\item[b:] 
$m(\lambda) = (\lambda-\lambda_1)^3(\lambda-\lambda_2)$ 
and $c(\lambda)=(\lambda-\lambda_1)^3(\lambda-\lambda_2)$. We have
\begin{equation}\label{20}
T^a{}_b \sim \left[\begin{array}{ccc|c} 
\lambda_1 &1 &0 &0\\ 0 &\lambda_1& 1&0\\  0&0& \lambda_1 & 0\\ \hline 0&0&0&\lambda_2\\\end{array}\right].
\end{equation}
This is a generic type III stress-energy tensor.\\
This tensor \emph{does not occur} classically in nature, and does not even seem to occur semi-classically.

\end{description}

\item[III:] Three distinct eigenvalues. We now have $m(\lambda) = (\lambda-\lambda_1)^2(\lambda-\lambda_2)(\lambda-\lambda_3)$ and $c(\lambda)=(\lambda-\lambda_1)^2(\lambda-\lambda_2)(\lambda-\lambda_3)$. Then
\begin{equation}\label{21}
T^a{}_b \sim \left[\begin{array}{cc|cc} 
\lambda_1 &1 &0 &0\\ 0 &\lambda_1& 0 &0\\ \hline 0&0& \lambda_2 & 0\\0&0&0&\lambda_3\\ \end{array}\right].
\end{equation}
This is a generic type II stress-energy tensor, (see references \cite{Peres, Bonnor} for specific examples of this kind of tensor).\\
Physically this corresponds, for example, to a \emph{null flux superimposed on a non-symmetric background}.

\item[IV:] Four distinct eigenvalues. Then $m(\lambda) = (\lambda-\lambda_1)(\lambda-\lambda_2)(\lambda-\lambda_3)(\lambda-\lambda_4)$ and $c=(\lambda-\lambda_1)(\lambda-\lambda_2)(\lambda-\lambda_3)(\lambda-\lambda_4)$. 
We have:
\begin{equation}
T^a{}_b \sim \left[\begin{array}{cccc} 
\lambda_1 &0 &0 &0\\ 0 &\lambda_2& 0 &0\\  0&0& \lambda_3 & 0\\0&0&0&\lambda_4\\ \end{array}\right].
\end{equation}
This is either generic type I, if all $\lambda_i$ are real, or generic type IV, if there are two complex and two real eigenvalues.\footnote{We cannot have four complex eigenvalues since that would correspond to two $2\times2$ blocks of (1+1) dimensional type IV, implying a physically inappropriate (2+2) signature.}

\end{description}

It should be noted that in the stress-energy tensors given by (\ref{12}) (\ref{14}), (\ref{15}), (\ref{16}), (\ref{18}), (\ref{20}), and (\ref{21}) the non-diagonal Jordan block appears in the timelike direction. This is because $T_{ab}$ is a symmetric tensor, which implies that its spatial Euclidean block is diagonalizable and, therefore, at least one of the spacelike Jordan blocks of $T^a{}_b$ also is diagonalizable.

\subsection{Generalized Rainich conditions}\label{sec:Rc}
As is well known~\cite{1925,Misner:1957,Witten:1959}, (see also appendix~\ref{A:B}), for the electromagnetic field the squared stress-energy tensor is proportional to the identity. Specifically
\begin{equation}
(T^2)_a{}^b =  \left\{ {1\over 4} \{|\vec E|^2- |\vec B|^2\}^2 +\{\vec E\cdot\vec B\}^2 \right\} \delta_a{}^b.  
\end{equation}
As for electromagnetism one has $T=\tr(T)=0$, in general relativity this implies
\begin{equation}
 (R^2)_a{}^b \propto \delta_a{}^b,
\end{equation}
which is the algebraic Rainich condition.
The new classification that we have presented above allows us to show that this is just a particular case of the more general relation that can be obtained for degree 2 stress-energy tensors.

\paragraph{Degree 1:} Then $T^a{}_b$ necessarily has only 1 distinct eigenvalue $\lambda_*$.
So in this case $m(T^\bullet{}_\bullet)=T^\bullet{}_\bullet-\lambda_* \,\delta^\bullet{}_\bullet=0$. For the stress-energy tensor
$T^a{}_b= {1\over4} T\,\delta^a{}_b$
with $T={\rm tr}(T^\bullet{}_\bullet)=T^a{}_a$. So, in general relativity we have
$ R^a{}_b= \frac{1}{4}R\,\delta^a{}_b$.
That is, for a degree 1 stress-energy tensor the geometry necessarily is an Einstein space-time.

\paragraph{Degree 2:} If $m(\lambda)$ has degree 2, there are two sub-cases.
\begin{description}

\item[I:] $T^a{}_b$ has only 1 distinct eigenvalue $\lambda_*$, then $m(T^\bullet{}_\bullet)=(T^2)^\bullet{}_\bullet-2\lambda_*  T^\bullet{}_\bullet+\lambda_*^2 \delta^\bullet{}_\bullet=0$. This implies
\begin{equation}
 (T^2)^a{}_b=\frac{\tr(T)}{2} \, T^a{}_b-\frac{\tr(T)^2}{16}\, \delta^a{}_b\quad\Longrightarrow\quad {\rm tr}(T^2)=\frac{\tr(T)^2}{4}.
\end{equation}
Assuming the Einstein equations, we translate this into the geometric condition
\begin{equation}
 (R^2)^a{}_b=\frac{1}{2}\tr(R)\,R^a{}_b-\frac{1}{16}\tr(R)^2\,\delta^a{}_b\quad\Longrightarrow\quad {\rm tr}(R^2)=\frac{1}{4}\tr(R)^2.
\end{equation}

\item[II:] $T^a{}_b$ has 2 distinct eigenvalues. Then 
\begin{equation}
 (T^2)^a{}_b=(\lambda_1+\lambda_2) T^a{}_b-\lambda_1\lambda_2 \delta^a{}_b,
\end{equation}
which leads to the geometrical condition
\begin{equation}
 (R^2)^a{}_b=\alpha R^a{}_b-\beta \delta^a{}_b,
\end{equation}
with $\alpha=\tr(R)+(\lambda_1+\lambda_2)\kappa$ and $\beta={1\over4}\left[\tr(R)^2+2\tr(R)(\lambda_1+\lambda_2)\kappa+4\lambda_1\lambda_2\kappa^2\right]$, and $\kappa=8\pi G$.\\
Note that for the particular case $\lambda_1=-\lambda_2$, we will have
\begin{equation}
\hspace{-1cm}
 (T^2)^a{}_b =\lambda_1^2 \delta^a{}_b,\quad\Longrightarrow\quad (R^2)^a{}_b=\tr(R) R^a{}_b-\frac{1}{4}\left[\tr(R)^2-{\rm tr}(R^2)\right] \delta^a{}_b.
\end{equation}
Noting that in this specific case $\tr(T)=\tr(R)=0$, this reduces to the Rainich condition for classical electromagnetism $(R^2)^a{}_b = {1\over4}\tr(R^2)\, \delta^a{}_b$. If we do not impose this specific relationship between $\lambda_1$ and $\lambda_2$ then the generalized Rainich condition $(R^2)^a{}_b=\alpha R^a{}_b-\beta \delta^a{}_b$ is appropriate for geometrizing both perfect fluid sources and/or scalar field sources~\cite{Krongos:2015}. 
\end{description}

\noindent
For degree 2 a nice result is to note that $T^2 = A \,T + B\,I$ implies $T^3 = A \,T^2 + B\,T$, so that taking traces
\begin{equation}
\tr(T^2) = A\, \tr(T) + 4\, B; \qquad\qquad \tr(T^3) = A\, \tr(T^2) + B\,\tr(T).
\end{equation}
These simultaneous linear equations can be solved for $A$ and $B$, with the general result that for degree 2 we have the explicit expression
\begin{equation}
\hspace{-0.5cm}
(T^2)^a{}_b =  \left\{ \tr(T)\, \tr(T^2) - 4 \tr(T^3)  \over  \tr(T)^2 - 4 \tr(T^2) \right\}  T^a{}_b    -
 \left\{ \tr(T^2)^2 - \tr(T) \,\tr(T^3)  \over  \tr(T)^2 - 4 \tr(T^2) \right\} \delta^a{}_b.
\end{equation}
If we work with the  traceless piece of the stress-energy $\hat T^a{}_b = T^a{}_b - {1\over 4} T \, \delta^a_b$, then (noting that the distribution of blocks in the Jordan normal form, and so the degree of the minimal polynomial, is left unchanged when the tensor is shifted by a multiple of the identity),  this simplifies to
\begin{equation}
(\hat T^2)^a{}_b =  \left\{ \tr(\hat T^3)  \over  \tr(\hat T^2) \right\}  \hat T^a{}_b  
 + {1\over 4} \tr(\hat T^2)\, \delta^a{}_b.
\end{equation}
While this is not precisely the usual Rainich condition it is remarkably close. (The classical electromagnetic Rainich condition corresponds to $\tr(T^3)=0=\tr(T)$.) Working at the level of geometry, since the Einstein equation relates the stress-energy to the Ricci tensor shifted by a multiple of the identity, for degree 2 the equivalent statement for the Ricci tensor is
\begin{equation}\label{R2}
\hspace{-0.5cm}
(R^2)^a{}_b =  \left\{ \tr(R)\, \tr(R^2) - 4 \tr(R^3)  \over  \tr(R)^2 - 4 \tr(R^2) \right\}  R^a{}_b    -
 \left\{ \tr(R^2)^2 - \tr(R) \,\tr(R^3)  \over  \tr(R)^2 - 4 \tr(R^2) \right\} \delta^a{}_b.
\end{equation}
(A \emph{massless minimally coupled scalar field} corresponds to  $R_{ab}= \nabla_a \phi \; \nabla_b\phi$, so that $\tr(R^m) =  (\nabla\phi\cdot\nabla\phi)^m = \tr(R)^m$, implying $R^2 = \tr(R)\, R$.)
For the traceless part of the Ricci tensor, $\hat R^a{}_b = R^a{}_b - {1\over 4} R\, \delta^a_b$, the discussion above  simplifies to
\begin{equation}\label{R22}
(\hat R^2)^a{}_b =  \left\{ \tr(\hat R^3)  \over  \tr(\hat R^2) \right\}  \hat R^a{}_b  
 + {1\over 4} \tr(\hat R^2)\, \delta^a{}_b.
\end{equation}
Note that both equation (\ref{R2}) and equation (\ref{R22}) are purely geometric conditions;  they therefore generalize the usual Rainich condition for any degree 2 stress-energy tensor. We can easily recover the usual Rainich condition directly from equation (\ref{R2}) when $\tr(R^3)=\tr(R)=0$.

\paragraph{Degree 3:} Let us treat all of the sub-cases for degree 3 together. We have 
\begin{eqnarray}
\hspace{-0.5cm}
(T^3)^a{}_b=(\lambda_1+\lambda_2+\lambda_3)\,(T^2)^a{}_b
-(\lambda_1\lambda_2+\lambda_1\lambda_3+\lambda_2\lambda_3)\,T^a{}_b+\lambda_1\lambda_2\lambda_3\,\delta^a{}_b,
\end{eqnarray}
with $\lambda_1=\lambda_2=\lambda_3$ for case I, $\lambda_1=\lambda_2\neq\lambda_3$ for case II, and $\lambda_1\neq\lambda_2\neq\lambda_3$ for case III.
We then obtain the following geometric equation
\begin{equation}
(R^3)^a{}_b=\alpha_2\,(R^2)^a{}_b-\alpha_1\, R^a{}_b+\alpha_0\,\delta^a{}_b,
\end{equation}
where
\begin{eqnarray}
\alpha_2&=&\frac{3}{2}\tr(R)+(\lambda_1+\lambda_2+\lambda_3)\kappa,\\
\alpha_1&=&\frac{3}{4}\tr(R)^2+\frac{1}{4}\tr(R)^2(\lambda_1+\lambda_2+\lambda_3)\kappa+(\lambda_1\lambda_2+\lambda_1\lambda_3+\lambda_2\lambda_3)\kappa^2,\\
\alpha_0&=&\frac{1}{8}\tr(R)^3+\frac{1}{4}\tr(R)^2(\lambda_1+\lambda_2+\lambda_3)\kappa+\frac{1}{2}\tr(R)(\lambda_1\lambda_2+\lambda_1\lambda_3+\lambda_2\lambda_3)\kappa^2\nonumber\\
&+&\lambda_1\lambda_2\lambda_3\kappa^3.\;\;\;
\end{eqnarray}

For degree 3 an explicit result in terms of traces of powers of the stress-energy is possible but is unfortunately somewhat unedifying. Noting that $T^3 = A \,T^2 + B\,T + C\, I$ implies both $T^4 = A \,T^3 + B\,T^2 + C\, T$ and $T^5 = A \,T^4 + B\,T^3 + C\, T^2$, taking traces yields
\begin{equation}
\hspace{-1.5cm}
\tr(T^3) = A\, \tr(T^2) + B\, \tr(T) + 4\, C; \quad \tr(T^4) = A\, \tr(T^3) + B\, \tr(T^2) + C \,\tr(T); 
\end{equation}
and
\begin{equation}
 \tr(T^5) = A\, \tr(T^4) + B\, \tr(T^3) + C\,\tr(T^2).
\end{equation}
These simultaneous linear equations can be solved for $A$, $B$, and $C$, resulting in an explicit but ugly expression for degree 3 that does not seem worth writing out.
Note that, analogously with the previous case, once we have an expression for $T^3$ in terms of lower powers of $T$ and $\tr(T^m)$ with $m\leq3$, we can consider a shift to find an expression of $R^3$ in terms of lower powers of $R$ and $\tr(R^m)$ with $m\leq3$.

A more subtle construction is this: For degree 3 at least one eigenvalue $\lambda_*$ is doubled, (or even tripled or quadrupled), and corresponds to a spacelike eigenvector~$s_a$. Eliminate this spacelike eigenvector by defining
\begin{equation}
(T')^a{}_b = T^a{}_b - \lambda_*\; s^a s_b. 
\end{equation}
The tensor $T'$ is now a singular matrix, and has only 3 eigenvalues corresponding to those occurring in the minimal polynomial $m(\lambda)$ of $T$. Now we can write
\begin{equation}
(T^3)^a{}_b =  \tr(T') \, (T^2)^a{}_b + {1\over2}\left\{\tr([T']^2) - \tr(T')^2\right\}\,T^a{}_b + \pdet(T')\, \delta^a{}_b.
\end{equation}
Here $\pdet(T')$ is the pseudo-determinant, the product over non-zero eigenvalues. This expression is simple and evocative, but somewhat implicit.

\paragraph{Degree 4:} Analogously, for degree 4 we have
\begin{eqnarray}
(T^4)^a{}_b&=&(\lambda_1+\lambda_2+\lambda_3+\lambda_4)\,(T^3)^a{}_b\nonumber\\
&&-(\lambda_1\lambda_2+\lambda_1\lambda_3+\lambda_2\lambda_3+\lambda_1\lambda_4+\lambda_2\lambda_4+\lambda_3\lambda_4)\,(T^2)^a{}_b\nonumber\\
&&+(\lambda_1\lambda_2\lambda_3+\lambda_1\lambda_2\lambda_4+\lambda_1\lambda_3\lambda_4+\lambda_2\lambda_3\lambda_4)\,T^a{}_b
-\lambda_1\lambda_2\lambda_3\lambda_4\,\delta^a{}_b.
\end{eqnarray}
(Here $\lambda_1=\lambda_2=\lambda_3\neq\lambda_4$ for case II, $\lambda_1=\lambda_2\neq\lambda_3\neq\lambda_4$ for case III, while $\lambda_1\neq\lambda_2\neq\lambda_3\neq\lambda_4$ for case IV.)

We can now easily re-express this in terms of the elementary symmetric polynomials as
\begin{eqnarray}
\hspace{-35pt}
(T^4)^a{}_b&=&e_1(T)\,(T^3)^a{}_b
-e_2(T)\,(T^2)^a{}_b
+e_3(T)\,T^a{}_b
-e_4(T)\,\delta^a{}_b.
\end{eqnarray}
Because we are in 4 dimensions, the general explicit formula for the third symmetric polynomial, $e_3(T) =\frac{1}{6}\left[(\tr(T)^3-3\tr(T)\tr(T^2)+2\tr(T^3))\right]$, can be more compactly rewritten in terms of the cofactor matrix,\footnote{If a matrix $X$ is nonsingular, then the cofactor matrix is $\cof(X) = \det(X) \, (X^{-1})^T$, but the cofactor matrix continues to make sense even if the matrix is singular.} $e_3(T) = \tr[\cof(X)]$, 
while $e_4(T)$ reduces to $\det(T)$. Thence
\begin{eqnarray}
\hspace{-35pt}
(T^4)^a{}_b&=&\tr(T)\,(T^3)^a{}_b
+{1\over2}\{\tr(T^2) - \tr(T)^2\}\,(T^2)^a{}_b\nonumber\\
&+& \{\tr[\cof(T)]\}\,T^a{}_b
-\det(T)\,\delta^a{}_b.
\end{eqnarray}

Similarly we obtain the geometric relation
\begin{eqnarray}
(R^4)^a{}_b&=&\tr(R)\,(R^3)^a{}_b
+{1\over2}\{\tr(R^2) - \tr(R)^2\}\,(R^2)^a{}_b\nonumber\\
&+& 
\{\tr[\cof(R)]\}\,R^a{}_b-\det(R)\,\delta^a{}_b.
\end{eqnarray}
By considering the traceless part of the stress-energy and Ricci tensors we can write
\begin{eqnarray}
(\hat T^4)^a{}_b&=&
+{1\over2}\{\tr(\hat T^2) \}\,(T^2)^a{}_b
+ 
\{\tr[\cof(\hat T)]\}\,\hat T^a{}_b
-\det(\hat T)\,\delta^a{}_b,\;\;
\end{eqnarray}
and
\begin{eqnarray}
(\hat R^4)^a{}_b&=&
+{1\over2}\{\tr(\hat R^2) \}\,(\hat R^2)^a{}_b + 
\{\tr[\cof(\hat R)]\}\,\hat R^a{}_b
-\det(\hat R)\,\delta^a{}_b.
\end{eqnarray}
The geometric conditions presented in this section show that the effect of \emph{any} stress-energy tensor can be described considering expressions written just with invariants of the Ricci curvature (traces of powers, the determinant, the trace of the cofactor matrix). 
Hence, any physical  acceptable geometry (that is, generated by a reasonable stress-energy tensor) should satisfy one of the generalized (algebraic) Rainich conditions that we have obtained.
We do not consider in this paper the extension of the differential Rainich equation, related with the dynamics of the source of the curvature.

\subsection{Applications to the energy conditions}\label{sec:EC}
The relations between the different powers of the stress-energy tensor presented in the previous sections allow us in some cases to extract  information regarding relations with the energy conditions,
at least for degrees 1 and 2. For instance

\paragraph{Degree 1:} As we have $T^a{}_b= {1\over4} T\,\delta^a{}_b$, then: 
\begin{itemize}
\item The WEC is satisfied if and only if the trace energy condition (TEC) is fulfilled ($T\leq0$). \emph{This corresponds to positive vacuum energy, positive cosmological constant.}
\item A minimum requirement for the DEC to be satisfied is that the TEC is fulfilled.
\item The SEC, that is $V_a\,[T^a{}_b-{1\over2} \tr(T)\,\delta^a{}_b]V^b\geq0$, is satisfied if and only if the TEC is violated. Therefore, the WEC and the SEC cannot be simultaneously satisfied.
\end{itemize}

\paragraph{Degree 2:} For degree 2 the condition $T^2=A\,T+B\,I$ implies 
\begin{equation}
(T^2)_{ab} V^a V^b =  A\,T_{ab} V^a V^b + B\, (g_{ab} V^aV^b);  \quad \tr(T^2) = A\,\tr(T)+4B.
\end{equation}
Consequently, in degree 2, ``quadratic''  energy conditions, such as the DEC, FEC, and TOSEC, automatically reduce to linear conditions on the stress-energy. We have two cases.
\begin{description}
\item[I:] $T^a{}_b$ has only 1 distinct eigenvalue $\lambda_*$: $(T^2)^a{}_b=\frac{\tr(T)}{2} \, T^a{}_b-\frac{\tr(T)^2}{16}\delta^a{}_b$. This implies (see definitions in references~\cite{Martin-Moruno:2013a,Martin-Moruno:2013b,Martin-Moruno:2015,LNP}):
\begin{itemize}
\item ${\rm tr}(T^2) = {1\over4} T^2 \geq0$, so TOSEC is satisfied.
\item If the WEC is satisfied, then the TEC ($T\leq0$) is a necessary requirement for the FEC to be fulfilled.
\end{itemize}

\item[II:] $T^a{}_b$ has 2 distinct eigenvalues $\lambda_1$ and $\lambda_2$: $(T^2)^a{}_b=(\lambda_1+\lambda_2) \,T^a{}_b-\lambda_1\lambda_2\, \delta^a{}_b$.
\begin{itemize}
\item If we want to have any hope of the FEC and WEC to be simultaneously satisfied, at least one of the eigenvalues has to be negative.
\item If the NEC is satisfied, a necessary requirement for the FEC to be fulfilled is that at least one of the eigenvalues has to be negative.
\item If the TEC is satisfied, at least one of the eigenvalues has to be negative for the TOSEC to be fulfilled.
\item For the particular case $\lambda_1=-\lambda_2$, we will have $(T^2)^a{}_b =\lambda_1^2 \, \delta^a{}_b$. Then the FEC is satisfied.
\end{itemize}
\end{description}
Since energy conditions (as presently defined) entail the consideration of quantities that are linear or quadratic in the stress-energy tensor, we see that although interesting relations between inequalities may be found for stress-energy tensors of degrees 3 and 4, they will generically not relate just the energy conditions.

\section{Discussion and conclusions}\label{sec:d}

So what have we learned from this exercise? Mathematically the (mixed component) stress-energy tensor forms a closed algebraic field of degree at most 4 over the real numbers. Algebraically, in (3+1) dimensions there will always be some exponent $1\leq N\leq4$ such that
\begin{equation}
(T^N)^\bullet{}_\bullet = \sum_{i=0}^{N-1}  k_i \; (T^i)^\bullet{}_\bullet.
\end{equation}
Physically, powers of stress tensors close in on themselves rather rapidly. Even for the worst behaved stress-energy tensor in (3+1) dimensions the 4th power is always expressible in terms of lower powers. Simple (and physically attractive)  stress-energy tensors often exhibit this behaviour even at 2nd order. 
As (currently defined) point-like energy conditions entail the consideration of terms linear or quadratic in the stress-energy tensor, some relations between the fulfillment of some of those energy conditions can be found for stress-energy tensors of degrees 1 and 2.
On the other hand, for each degree of the classification based on the minimal polynomials, one can write a purely geometric expression for the curvature of the corresponding spacetime. The resulting expressions can be interpreted as generalized Rainich conditions that will \emph{always} be satisfied.
This construction gives us an alternative way of classifying stress-energy tensors, often providing a refinement of the usual classical and/or semi-classical energy conditions~\cite{Hell,Martin-Moruno:2013a,Martin-Moruno:2013b,Martin-Moruno:2015,LNP,Visser:1994,Visser:1997,Balakrishnan:2017,Fu:2017,Fu:2016, Koeller:2015, Bousso:2015} and/or the Hawking--Ellis (Segr\'e--Pleba\'nski) classification~\cite{Hell,Plebanski:1964,LNP}. 


\section*{Acknowledgments}
PMM acknowledges financial support from the projects FIS2014-52837-P (Spanish MINECO) and FIS2016-78859-P (AEI/FEDER, UE).
MV acknowledges financial support via the Marsden Fund administered by the Royal Society of New Zealand.

\appendix
\addappheadtotoc
\appendixpage

\section{Hawking--Ellis (Segr\'e--Pleba\'nski) classification}
\label{A:A}
In this appendix we will be using $\sim_{\hbox{\tiny L}}$ to denote similarity under Lorentz transformations; whereas $\sim$ will be used to denote similarity under generic non-singular transformations (used to get the Jordan normal form).
Similarity properties under Lorentz transformations are sketched in reference~\cite{Hell} and discussed more extensively in reference~\cite{LNP}. The Jordan normal form is mathematically convenient~\cite{HJ1,HJ2} but often more subtle to interpret physically --- similarity under generic non-singular transformations does not have a direct clean physical interpretation.
\begin{description}
\item[type I]:

\begin{equation}
\hspace{-1cm}
T^{ab} \sim_{\hbox{\tiny L}} \left[\begin{array}{c|ccc} 
\rho&0 &0 &0\\ \hline 0 &p_1& 0 &0\\  0&0& p_2 & 0\\0&0&0&p_3\\\end{array}\right];
\qquad
T^a{}_b \sim\left[\begin{array}{c|ccc} 
-\rho&0 &0 &0\\ \hline 0 &p_1& 0 &0\\  0&0& p_2 & 0\\0&0&0&p_3\\\end{array}\right].
\end{equation}
 Eigenvalues: $\{-\rho,p_1,p_2,p_3\}$.

\item[type II]:

\begin{equation}
\hspace{-2cm}
T^{ab} \sim_{\hbox{\tiny L}} \left[\begin{array}{cc|cc} 
\mu+f&f &0 &0\\ f &-\mu+f& 0 &0\\ \hline 0&0& p_2 & 0\\0&0&0&p_3\\\end{array}\right];
\quad
T^a{}_b \sim\left[\begin{array}{cc|cc} 
-\mu&1 &0 &0\\ 0 &-\mu& 0 &0\\ \hline 0&0& p_2 & 0\\0&0&0&p_3\\\end{array}\right].
\end{equation}
Eigenvalues: $\{-\mu,-\mu,p_2,p_3\}$.

\item[type III]:
\begin{eqnarray}
T^{ab}
\sim_{\hbox{\tiny L}} \left[\begin{array}{ccc|c} 
\rho&f &0 &0\\ f &-\rho& f &0\\  0&f& -\rho & 0\\   \hline 0&0&0&p_3\\\end{array}\right]
\sim_{\hbox{\tiny L}} \left[\begin{array}{ccc|c} 
\rho&0 &f &0\\ 0 &-\rho& f &0\\  f&f& -\rho & 0\\   \hline 0&0&0&p_3\\\end{array}\right];\nonumber\\
\qquad\qquad\quad
 T^a{}_b \sim\left[\begin{array}{ccc|c} 
-\rho&1 &0 &0\\ 0 &-\rho& 1 &0\\  0&0& -\rho & 0\\ \hline 0&0&0&p_3\\\end{array}\right].
\end{eqnarray}
Eigenvalues: $\{-\rho,-\rho,-\rho,p_3\}$.

\item[type IV]:
\begin{equation}
\hspace{-2cm}
T^{ab} \sim_{\hbox{\tiny L}} \left[\begin{array}{cc|cc} 
\rho&f &0 &0\\ f &-\rho& 0 &0\\ \hline 0&0& p_2 & 0\\0&0&0&p_3\\\end{array}\right];
\qquad
T^a{}_b \sim\left[\begin{array}{cc|cc} 
-\rho+if&0 &0 &0\\ 0 &-\rho-if& 0 &0\\ \hline 0&0& p_2 & 0\\0&0&0&p_3\\\end{array}\right].
\end{equation}
Eigenvalues: $\{-\rho+if ,-\rho-if,p_2,p_3\}$.

\item[Hawking--Ellis in (1+1) dimensions:]
It is sometimes useful to explicitly consider the restriction of the Hawking--Ellis classification to (1+1) dimensions,  where type III  does not exist, but types I, II, and IV simplify to:
\begin{equation}
T^{ab} \sim_{\hbox{\tiny L}} \left[\begin{array}{c|c} 
\rho&0 \\ \hline 0 &p\end{array}\right];
\quad
T^a{}_b \sim\left[\begin{array}{c|c} 
-\rho&0\\ \hline 0 &p\end{array}\right];
\quad
\hbox{eigenvalues: } \{-\rho,p\}.
\end{equation}
\begin{equation}
\hspace{-2cm}
T^{ab} \sim_{\hbox{\tiny L}} \left[\begin{array}{c|c} 
\mu+f&f \\ \hline f &-\mu+f\end{array}\right];
\,\,
T^a{}_b \sim\left[\begin{array}{c|c} 
-\mu&1\\ \hline 0 &-\mu\end{array}\right];
\,\,
\hbox{eigenvalues: } \{-\mu,-\mu\}.
\end{equation}
\begin{equation}
\hspace{-2cm}
T^{ab} \sim_{\hbox{\tiny L}} \left[\begin{array}{c|c} 
\rho&f \\ \hline f &-\rho\end{array}\right];
\,\,
T^a{}_b \sim\left[\begin{array}{c|c} 
-\rho+if&0\\ \hline 0 &-\rho-if\end{array}\right];
\,\,
\hbox{eigenvalues: } \{-\rho\pm if\}.
\end{equation}
These $2\times2$ blocks are sometimes useful when building up a 4 dimensional analysis.

\end{description}

\section{The classic Rainich result for electromagnetism}
\label{A:B}
\def\tr{{\mathrm{tr}}}
The electromagnetic stress-energy tensor is
\begin{equation}
T_{ab}  = -F_{ac} g^{cd} F_{db} - {1\over 4} (F_{cd} F^{cd}) g_{ab}.
\end{equation}
Raising one index
\begin{equation}
T_a{}^b  = -(F^2)_a{}^b + {1\over 4} \tr(F^2) \delta_a{}^b.
\end{equation}
Then
\begin{equation}
(T^2)_a{}^b = (F^4)_a{}^b -{1\over2}  \tr(F^2) \,(F^2)_a{}^b + {1\over 16} \tr(F^2)^2\, \delta_a{}^b.
\end{equation}
A little algebra, using the antisymmetry of $F_{ab}$, now yields~\cite{1925,Misner:1957,Witten:1959}\footnote{{To see roughly why this works note that the antisymmetry of $F_{ab}$ implies
\[
(F^4)_a{}^b = A \,(F^2)_a{}^b  + B \, \delta_a{}^b = \bar A \, T_a{}^b + \bar B \, \delta_a{}^b.
\]
Furthermore $(F^2)_a{}^b = \bar C \, T_a{}^b + \bar D \, \delta_a{}^b$. Thence $(T^2)_a{}^b = \bar E \, T_a{}^b + \bar F \, \delta_a{}^b$. Taking the trace, $\bar F = {1\over4} \tr(T^2)$. 
But the coefficient $\bar E$ must be linear in $T^a{}_b$, so it is proportional to $\tr(T)$, which is zero. QED.}
}
\begin{equation}
(T^2)_a{}^b = {1\over 4} \tr(T^2)\, \delta_a{}^b.
\end{equation}
In terms of the Lorentz invariants $|\vec E|^2- |\vec B|^2$ and $\vec E\cdot\vec B$ this reads
\begin{equation}
(T^2)_a{}^b =  \left\{ {1\over 4} \{|\vec E|^2- |\vec B|^2\}^2 +\{\vec E\cdot\vec B\}^2 \right\} \delta_a{}^b.  
\end{equation}

\section{ Rainich classification in $1+1$ dimensions }
\label{A:C}
$(1+1)$-dimensional scenarios are usually considered as toy models which can provide us with information of physical interest. Stress-energy tensors for these scenarios can be only of type I, II, and IV according to the Hawking--Ellis classification. In this appendix we consider the classification introduced in this paper in terms of the minimal polynomial. In $1+1$ dimensions the stress-energy tensors can be, therefore, classified as follows:

\paragraph{Degree 1:} The only possibility is $m(\lambda)=(\lambda-\lambda_*)$ and $c(\lambda)=(\lambda-\lambda_*)^2$. That is, we have only one eigenvalue that, therefore, is real.
\begin{equation}
T^a{}_b \sim \left[\begin{array}{cc} 
\lambda_* & 0 \\ 0 & \lambda_* \\\end{array}\right].
\end{equation}
This is a special case of type I with $\lambda_*=-\rho=p$.\\
For this case we have that $m(T^\bullet{}_\bullet)=T^\bullet{}_\bullet-\lambda_*\delta^\bullet{}_\bullet=0$. Then
\begin{equation}
T^a{}_b=\frac{1}{2}\tr(T)\,\delta^a{}_b,
\end{equation}
which describes vacuum energy.

\paragraph{Degree 2:} There are two subcases:
\begin{description}
\item[I:] There is only one distinct eigenvalue, which has to be real. That is, $m(\lambda)=(\lambda-\lambda_*)^2$ and $c(\lambda)=(\lambda-\lambda_*)^2$.
So
\begin{equation}
T^a{}_b \sim \left[\begin{array}{cc} 
\lambda_* & 1 \\ 0 & \lambda_* \\\end{array}\right].
\end{equation}
This is a type II stress-energy tensor with $\lambda_*=-\mu$.

\item[II:] There are two distinct eigenvalues. So, $m(\lambda)=(\lambda-\lambda_1)(\lambda-\lambda_2)$ and $c(\lambda)=(\lambda-\lambda_1)(\lambda-\lambda_2)$, with
\begin{equation}
T^a{}_b \sim \left[\begin{array}{cc} 
\lambda_1 & 0 \\ 0 & \lambda_2 \\\end{array}\right].
\end{equation}
This is a generic type I, if the eigenvalues are real, and a type IV, if they are complex.

For both subcases we have
\begin{equation}
(T^2)^a{}_b=\tr(T) \,T^a{}_b-{\rm det}(T)\,\delta^a{}_b.
\end{equation}
\end{description}
In counterpoint we note that the Einstein equations are meaningless in 1+1 dimensions, 
$R^a{}_b={1\over 2} R \, \delta^a{}_b$ identically.

\section*{References}

  
%
\end{document}